\documentclass[twocolumn,showpacs,preprintnumbers,amsmath,amssymb,aps,prd,nofootinbib,superscriptaddress,eqsecnum]{revtex4}
\usepackage{makeidx}
\usepackage[dvipdfmx]{graphicx}
\usepackage[all]{xy}
\usepackage{feynmp}
\usepackage{color}
\usepackage{amsmath,amssymb,bm}

\renewcommand{\eqref}[1]{Eq.~(\ref{#1})}
\newcommand{\bee}[1]{\begin{align}#1\end{align}}
\newcommand{\mt}[1]{\begin{matrix} #1 \end{matrix}}
\newcommand{\m}[1]{{\rm #1}}
\newcommand{\nn}{\nonumber}
\newcommand{\ra}{\rightarrow}
\newcommand{\figref}[1]{Fig.~\ref{#1}}
\newcommand{\mc}{\mathcal}
\newcommand{\tr}{\mathrm{Tr}}
\newcommand{\pd}{\partial}
\newcommand{\braa}[1]{\left( #1 \right)}

\newcommand{\brac}[1]{\left[ #1 \right]}

\newcommand{\brae}[1]{\langle #1 \rangle}

\begin{document}

\title{Equation of state in the pion condensation phase in the asymmetric nuclear matter using a holographic QCD model}

\author{Hiroki Nishihara\footnote{h248ra@hken.phys.nagoya-u.ac.jp}}
\author{Masayasu Harada\footnote{harada@hken.phys.nagoya-u.ac.jp}}
\affiliation{
Department of Physics, Nagoya University, Nagoya 464-8602, Japan
}

\date{\today}

\def\theequation{\thesection.\arabic{equation}}
\makeatother

\begin{abstract}
We study the  asymmetric nuclear matter using a holographic QCD model by introducing a baryonic charge in the infrared boundary.
We first show that, in the normal hadron phase, the predicted values of the symmetry energy and it's slope parameter are comparable with the empirical values.
We find that the phase transition from the normal phase to the pion condensation phase is delayed compared with the pure mesonic matter: The critical chemical potential is larger than the pion mass which is obtained for the pure mesonic matter.
We also show that, in the pion condensation phase, the pion contribution to the isospin number density increases with the chemical potential, while the baryonic contribution is almost constant.  Furthermore, the value of chiral condensation implies that the enhancement of the chiral symmetry breaking occurs in the asymmetric nuclear matter as in the pure mesonic matter.
We also give a discussion on how to understand the delay in terms of the 4-dimensional chiral Lagrangian including the rho and omega mesons based on the hidden local symmetry.
\end{abstract}


\maketitle

\section{Introduction}
\label{sec:Introduction}

It is expected that investigation of the hadron physics in extreme conditions  will give a clue for  our understanding of QCD (Quantum Chromodynamics).
In particular studying asymmetric nuclear matter is also important to derive the equation of state inside neutron stars~\cite{Lattimer:2006xb}, which will give a clue to understand the recently found very heavy neutron star~\cite{Demorest:2010bx,Antoniadis:2013pzd}.

We often draw the QCD phase diagram on the plane of temperature $T$ and the baryon chemical potential $\mu_B$~\cite{Fukushima:2010bq,Fukushima:2013rx}. It is expected that various phases exist on the plane $(T,  \mu_B)$ of the phase diagram: e.g. the quark-gluon plasma phase and the color superconducting phase. Similarly, 
finite isospin chemical potential $\mu_I$ provide a rich phase structure
which includes the pion condensation phase. 
There are many works studying the phase diagram at $\mu_I\neq0$.
In particular, the pion condensation phase transition on the plane $(\mu_B, \mu_I)$ are studied by introducing $\mu_I$ together with $\mu_B$ in the Nambu-Jona-Lasinio (NJL) model~\cite{Sasaki:2010jz,Zhang:2006gu,Toublan:2003tt,
Barducci:2004tt,He:2005nk,Mu:2010zz,He:2006tn,He:2005tf,Zhang:2013oia,Andersen:2007qv} and holographic QCD models~\cite{Lee:2013oya,Parnachev:2007bc}
and so on~\cite{Klein:2003fy,Nishida:2003fb,Toublan:2004ks,Abuki:2013vwa}.

As a first step to study the rich phase structure, it is interesting to study the
phase transition from the normal hadron phase to 
the pion condensation phase  
together with the equation of state 
in the pion condensation phase on the plane $(\mu_B, \mu_I)$.
References~\cite{Sasaki:2010jz,Zhang:2006gu,He:2005nk,Barducci:2004tt,Toublan:2003tt,Mu:2010zz} show the 
$T$ \textendash $\mu_B$ \textendash $\mu_I$ phase diagram via the NJL model, in which the dependence of the isospin density on the isospin chemical potential is shown only for $T=\mu_B=0$.
On the other hand, by using holographic QCD models~\cite{Parnachev:2007bc,Lee:2013oya}, the pion condensation phase transition is discussed. 
In Ref.~\cite{Parnachev:2007bc}, they draw the phase diagram on the plane $(\mu_B, \mu_I)$ in the Sakai-Sugimoto model.
Reference~\cite{Lee:2013oya} also studies stability of the normal hadron phase at finite isospin density by introducing the baryon charge as the Reissner-Nordstr\"{o}m (RN) blackhole charge in a hard wall holographic QCD model.
However, the equation of state in the pion condensation phase is not discussed in these works.

In the previous work~\cite{Nishihara:2014nva}, we studied the pion condensation in the pure mesonic matter using a holographic QCD model by introducing the isospin chemical potential as a UV boundary value of the gauge field.
We showed that the phase transition from the normal hadron phase to the pion condensation phase is of the second order and the critical value of the isospin chemical potential is equal to the pion mass, consistently with the chiral Lagrangian analysis~\cite{Son:2000xc}.

In Ref.~\cite{Nishihara:2014nva}, we studied the $\mu_I$-dependence of the chiral condensate defined by
$\tilde{\sigma} \equiv \sqrt{ \langle \sigma \rangle^2 + \langle \pi^a \rangle^2 }$, and showed that, although the ``$\sigma$"-condensate decreases rapidly with the isospin chemical potential in the pion condensation phase, the $\pi$-condensate increases more rapidly.  As a result the chiral condensate $\tilde{\sigma}$ keeps increasing, which implies the enhancement of the chiral symmetry breaking in the pion condensation phase.
The symmetry structure for this is understood in the following way:
When the isospin chemical potential is introduced, the chiral symmetry $\mbox{SU}(2)_R \times \mbox{SU}(2)_L$ is explicitly broken to $\mbox{U}(1)_R^{(3)} \times \mbox{U}(1)_L^{(3)} = \mbox{U}(1)_V^{(3)} \times \mbox{U}(1)_A^{(3)}$, where the superscript $^{(3)}$ implies that the generator $T_3$ of SU(2) is used for the U(1) as $\exp[ i \theta_V T_3 ] \in \mbox{U}(1)_V^{(3)}$.
In the normal hadron phase the $\mbox{U}(1)_A^{(3)}$  is broken by the ``$\sigma$''-condensate spontaneously and the quark mass explicitly.
In the pion condensation phase, on the other hand, the $\mbox{U}(1)_V^{(3)}$ symmetry  is spontaneously broken by the $\pi$-condensate, which generates a massless Nambu-Goldstone boson.
Since both $\mbox{U}(1)_A^{(3)}$  and $\mbox{U}(1)_V^{(3)}$ are subgroups of the chiral  $\mbox{SU}(2)_R \times \mbox{SU}(2)_L$ symmetry, the above structure implies that the chiral symmetry is never restored in the mesonic matter with the isospin chemical potential, and actually the breaking is enhanced in the pion condensation phase.
We note that the above properties are obtained in the pure mesonic matter, so that it is interesting to ask whether they are changed by the existence of the nucleon in the matter.

In this paper, we adopt a simple way for introducing the baryonic sources: We include a point-like nucleon source at the IR boundary coupling to the iso-triplet vector meson in the hard wall holographic QCD model as in Ref.~\cite{Kim:2007zm}, and studied the pion condensation in the asymmetric nuclear matter.
We will show that the phase transition from the normal hadron phase to the pion condensation phase is delayed in the asymmetric nuclear matter compared with the pure mesonic matter.
In other words, the critical chemical potential is larger than the pion mass.
On the other hand, the enhancement of the chiral symmetry breaking still occurs since the chiral condensate $\tilde{\sigma}$ keeps increasing with the isospin chemical potential.

This paper is organized as follows:
In section~\ref{sec:model}, we briefly review the holographic QCD model used in our analysis, and introduce the baryonic charge following Ref.~\cite{Kim:2007zm}.
Section~\ref{sec:delay} is devoted to the study of the symmetry energy and the pion mass in the normal hadron phase.
In section~\ref{sec:pion condensation},
we study the pion condensation phase and obtain the relation between the isospin chemical potential and the isospin number density as well as the chiral condensate.
In section~\ref{sec:HLS}, we make an analysis of the pion mass in the normal hadron phase using the four dimensional chiral model based on the hidden local symmetry~\cite{Bando:1987br,Harada:2003jx}.
We give a summary and discussions in section~\ref{sec:summary}.
We also show the equations of motion in appendix~\ref{app:EOM}.

\section{Model}
\label{sec:model}

In the present analysis, we employ 
a holographic QCD model given in Refs.~\cite{Erlich:2005qh,Da Rold:2005zs,DaRold:2005vr} for the mesonic part.
Then the mesonic action in the five dimensional space is given by
\bee{
S_5=S_X
+S^\m{BD}
}
where
\bee{
S_X=&\int d^4x \int_\epsilon^{z_m} dz
\nn\\&
\sqrt{g} \mathrm{Tr} \left\{ |DX|^2 -m_5^2|X|^2 -\frac{1}{4g_5^2}\left(F^2_L+F^2_R\right)\right\}
\ ,
\\
S^\m{BD}=&
-\int d^4x \int_\epsilon^{z_m} dz
\nn\\
&\sqrt{g} \ \mathrm{Tr} \left\{ \lambda z_m |X|^4-m^2 z_m |X|^2 \right\}\delta\left(z-z_m\right)
}
with $m_5^2=-3$.
The metric is written as
\bee{
ds^2&=a^2(z) \left(\eta_{\mu\nu}dx^\mu dx^\nu -dz^2\right) =g_{MN}dx^M dx^N
}
with 
\bee{
a(z)=\frac{1}{z}
\ ,
\label{metric}
}
where $z_m$ and $\epsilon$ are the IR-cutoff and UV-cutoff.
Here $N$ and $M$ run over 0,1,2,3,5 and $\eta_{\mu\nu}$ is the defined as the Mankowski metric: $\eta_{\mu\nu}={\rm diag} (1,-1,-1,-1)$.
\footnote{
Although there is a Chern-Simons term in addition, the term does not affect our result since we assume the rotational invariance in the present analysis.}

The model has the chiral symmetry U$(2)_L\times $U$(2)_R$ (=U$(1)_L\times $U$(1)_R\times$SU$(2)_L\times $SU$(2)_R$), under which
the fields transform in the following form:
\begin{eqnarray}
X &\rightarrow & X'= g_L X g_R^\dagger
\ ,
\label{X trans}\\
L_M &\rightarrow& {L'}_M = g_L L_M g_L^\dagger +i g_L \partial_M g_L^\dagger
\ ,
\\
R_M &\rightarrow& {R'}_M = g_R R_M g_R^\dagger +i g_R \partial_M g_R^\dagger
\label{L trans}
\end{eqnarray}
with $g_R \in \mbox{U}(2)_R$ and $g_L \in \mbox{U}(2)_L$.
The covariant derivative and the field strength are defined as
\begin{eqnarray}
D_M X&=&\partial_M X-iL_M X + i X R_M
\ ,
\\
F^L_{MN}&=&\partial_M L_N -\partial_N L_M - i\left[ L_M , L_N \right]
\label{strength}
\end{eqnarray}
and similar for $F^R_{MN}$.
These fields are 
parametrized as
\begin{eqnarray}
&&
L^I_M=\mathrm{Tr}\left[L_M\sigma^I\right] 
\ , 
\quad
R^I_M=\mathrm{Tr}\left[R_M\sigma^I\right] 
\ ,
\label{parameterize2}
\\
&&
V^I_M=\frac{R^I_M+L^I_M}{2} 
\ , 
\quad
A^I_M=\frac{R^I_M-L_M^I}{2}
\ ,
\label{parameterize4}
\\
&&
X=\frac{1}{2}\left(S^0 \sigma^0 +S^a \sigma^a  \right)e^{i\pi^b \sigma^b+i \eta}
\label{parameterize5}
\end{eqnarray}
where 
 $\sigma^I=(\sigma^0 ,\sigma^a)=(1 ,\sigma^a)$ and $\sigma^a$ are the Pauli matrices. 
In the following analysis we adopt the gauge $L_5=R_5=0$ and the IR boundary condition 
$\left. F^L_{5\mu}\right|_{z_m}=\left.F^R_{5\mu}\right|_{z_m}=0$.

Now, let us include the effects of the nucleon into the model.
Here we introduce baryonic sources for the quark number density $n_q$ and the baryonic contribution to the isospin number density, denoted by $n_I^{\rm {Baryon}}$, through the following term
\footnote{The sign of this term is uniquely determined from the definition of the chemical potential 
introduced 
in \eqref{isospin chemical}.}:
\bee{
S_\m{int}=\int d^4x \int_\epsilon^{z_m} dz~
\left[ V_0^0 n_q + V_0^3 n_I^{\rm {Baryon}} \right]
\delta\left(z-z_m+\delta z\right)
\label{Sint}
}
where $\delta z$ ($>0$) is an infinitesimal length and 
$V_0^0$ and $V_0^3$ are the gauge fields corresponding to the quark number density and isospin number density.
The baryon number density $n_B$ is defined as $n_B=n_q/N_c$.
We introduced the baryonic sources by the $\delta$-function having a peak near the IR boundary 
\cite{Kim:2007zm}, which doesn't modify the IR boundary conditions.

In the present analysis, 
we assume that the proton (neutron) does not appear as long as the proton  (neutron) chemical potential $\mu_p$ ($\mu_n$) is smaller than the mass of a nucleon, denoted by $m_N$.
Therefore 
our analysis will be done for the following three cases separately:
\bee{
&{\rm (i)}~~ -m_N \leq \mu_p < m_N \ , -m_N \leq \mu_n < m_N \ ,
\nn\\
&{\rm (ii)}~~~~~ m_N \leq \mu_p \ , ~~~~~~~~-m_N \leq \mu_n < m_N \ ,
\nn\\
&{\rm (iii)}~~~~m_N \leq \mu_p \ , ~~~~~~~~~~~~m_N \leq \mu_n  \ .
}
The proton and neutron chemical potential $\mu_p$ and $\mu_n$ are
 related with the isospin chemical potential $\mu_I$ and  the baryon chemical potential $\mu_B$ through $\mu_p=\mu_B+\mu_I/2$ and $\mu_n=\mu_B-\mu_I/2$.
The assumption implies $2n_I^{\rm Baryon}=n_B=0$ in Case-(i) and $2n_I^{\rm Baryon}=n_B=n_p$ in Case-(ii) because $n_I^{\rm Baryon}$ and $n_B$ are expressed as the difference between the proton density $n_p$ and the neutron density $n_n$ and the sum of them, respectively: $n_I^{\rm Baryon}=\frac{n_p-n_n}{2}\ , n_B=n_p+n_n$.
Case-(i) corresponds to the pure mesonic case which is studied in Ref.~\cite{Nishihara:2014nva}.
On the other hand, $n_I^{\rm Baryon}$ and $n_B$ are independent of the each other in Case-(iii).
We will show the results of our analysis in 
the Case-(ii) and the Case-(iii) 
to compare with the pure mesonic case.

We note that the 
four-dimensional part of the gauge symmetry is fixed when 
$S_{\rm int}$  is introduced. 
In other words, 
$S_{\rm int}$ is not invariant under the four-dimensional gauge transformation.
Then, we introduce the
quark number chemical potential $\mu_q$ and the isospin chemical potential $\mu_I$
as the UV boundary values of the time components of the gauge fields
as
\footnote{
The baryon number chemical potential $\mu_B$ is related to $\mu_q$ as $\mu_B\equiv N_c \mu_q$.
}
\bee{
\left. V_0^0\right|_\epsilon=\mu_q-c_{(0)}
\ , \quad
\left. V_0^3\right|_\epsilon=\mu_I-c_{(3)}
\ ,
\label{isospin chemical}
}
where 
the constants $c_{(0)}$ and $c_{(3)}$ are corresponding to the degree of freedom of the gauge transformation. 
In the next section, 
we will determine the values of $c_{(0)}$ and $c_{(3)}$ by the physical requirements
for the pion mass and the equation of state between the chemical potential and the density.

This holographic QCD model involves the following five parameters,
\bee{
g_5^2
\ ,~~
z_m
\ ,~~
m_q
\ ,~~
\lambda
\ ,~~
m^2
\ .
}
To match this model with QCD, the parameter $g_5^2$ is adjusted as~\cite{Erlich:2005qh} 
\bee{
\frac{1}{g_5^2}=\frac{N_c}{12\pi^2}
\ .
\label{g_5}
}
For the physical inputs to determine the parameters, we use
the pion mass $m_\pi=139.6$MeV, the pion decay constant $f_\pi=92.4$MeV, the $\rho$ meson mass $m_\rho=775.8$MeV, and the $a_0$ meson mass.
As in Ref.~\cite{Nishihara:2014nva}, we use the $a_0$ meson mass $m_{a_0} = 980\,$MeV as a reference value,
and see the dependence of our results on the scalar meson mass.
The values of the parameters corresponding to $m_{a_0} = 980\,$MeV are determined as
\bee{
z_m=1/(323 {\rm MeV})
\ ,~~
m_q=2.29 {\rm MeV}
\ ,~~
\nn\\
\lambda=4.4
\ ,~~
m^2=5.39
\ .
\label{parameters}
}

As in Ref.~\cite{Nishihara:2014nva},  we assume that the pion condensation phase has the rotational symmetry, $L_i=R_i=0$
\footnote{Note that we also take expectation values of operators made by $L_i$ and $R_i$ such as $\sum_i R_i R_i$, which are invariant under the rotational symmetry, vanish since our present analysis is of the leading order in the large $N_c$ expansion and the hadronic loop contributions are suppressed.
}, 
and the iso-triplet scalars do not condense, $S^a=0$.
Furthermore, we take $V_0^1=V_0^2=0$ and $A_0^3=\pi^3=0$ which form a set of solutions of the equation of motion (EOM) for these fields. 
Similarly, the set of the $\eta=0$ and $A_0^0=0$ satisfies the EOM and we take this solution.

The grand potential density $\Omega$ is given from the Lagrangian $\mc{L}$:
\bee{
\Omega =&-\int_\epsilon^{z_m} dz \mc{L}
\label{omega}
}
where the explicit form of the Lagrangian $\mc{L}$ is shown in \eqref{Lagrangian}.
One can derive the EOM for the vector field $V^0_0$ and $V^3_0$ from \eqref{omega}, 
\bee{
                 {} \pd_5\frac{a}{g_5^2}\pd_5 V_0^0
&=n_q \delta (z-z_m+\delta z)
\nn\ ,\\
                 {} \pd_5\frac{a}{g_5^2}\pd_5 V_0^3 
&-{} \frac{a^3\left(S^0\right)^2}{2}\left[2\sin^2 b ~V_0^3 
			+\theta\sin 2b \sin \zeta \right]
\nn \\ 
&~~~~~~~~~~~~~~~~~~~
=n_I^{\rm {Baryon}} \delta (z-z_m
+\delta z)
\ .
\label{EOM for v}
}
By parameterizing $V_0^0$ and $V_0^3$ as
\bee{
V_0^0(z)=&{\mu}_q-c_{(0)}+\varphi^0 (z)+g_5^2n_q \int_\epsilon^z d\tilde{z} \tilde{z} \theta (\tilde{z}-z_m+\delta z)
\ ,\nn\\
V_0^3(z)=&\mu_I-c_{(3)}+\varphi^3 (z)+g_5^2n_I^{\rm {Baryon}} \int_\epsilon^z d\tilde{z} \tilde{z} \theta (\tilde{z}-z_m+\delta z)
}
where $\theta$ is a step function,
\eqref{EOM for v} is rewritten as
\bee{
                  {} \pd_5\frac{a}{g_5^2}\pd_5 \varphi^0 
=&0
\ ,
\nn\\
                  {} \pd_5\frac{a}{g_5^2}\pd_5 \varphi^3 
=& \frac{a^3\left(S^0\right)^2}{2}\left[2\sin^2 b ~\braa{\mu_I-c_{(3)}+\varphi^3}\right.
\nn\\&
\left.
~~~~~~~~~~~~~~~~~~~
			+\theta\sin 2b \sin \zeta \right]
\ .
\label{eom for phi}
}
Here the boundary conditions are given by
\bee{
\left.V_0^{(0,3)} \right|_\epsilon=\mu_{(q,I)}-c_{(0,3)}
&\ ,~~
\left.\partial_5V_0^{(0,3)} \right|_{z_m}=0
\nn\\
\ra~~
\left.\varphi^{(0,3)} \right|_\epsilon=0
&\ ,~~
\left\{
\mt{
\left.\partial_5\varphi^{0} \right|_{z_m}=&-g_5^2z_m n_{q}
\\
\left.\partial_5\varphi^{3} \right|_{z_m}=&-g_5^2z_m n_I^{{\rm {Baryon}}}
}
\right.
\ .
\label{BD for V}
}

\section{Symmetry energy and delay of the phase transition}
\label{sec:delay}

In this section we first 
study the dependence of the pion mass on the isospin chemical potential $\mu_I$ in the normal hadron phase to show the delay of the pion condensation compared with the pure mesonic matter studied in Ref.~\cite{Nishihara:2014nva}.
Next, we investigate  the symmetry energy to check whether the present way to introduce the baryonic matter works well in the normal hadron phase by comparing our result with its empirical value.
For studying the hadron phase we set $b=0$ and $\theta=0$ in the equations of motion in Eqs.~(\ref{EOM for v}) and (\ref{eom for phi}).

In Case-(iii), we first derive the relation between the chemical potential $\mu_I$ and the isospin number density $n_I$.
For $b=\theta=0$, it is easy to solve the equation of motion (\ref{eom for phi}) with the boundary conditions in \eqref{BD for V} to have
\bee{
\varphi^3=&-\frac{g_5^2 n_I^{\rm {Baryon}}}{2} z^2
\ .
}
Substituting this solution into \eqref{omega}, 
we obtain
\bee{
\Omega 
\supset&
                  {} -\int_\epsilon^{z_m} dz~\frac{a}{2g_5^2}\left(\partial_5 V_0^3 \right)^2 
-n_I^{\rm {Baryon}} \left.V_0^3\right|_{z_m-\delta z}
\nn\\
=&
                  {} \frac{g_5^2 z_m^2}{4} \braa{n_I^{\rm {Baryon}}}^2
-\braa{\mu_I-c_{(3)}}n_I^{\rm {Baryon}} 
}
where $a=\frac{1}{z}$. Minimizing the $\Omega$ 
in terms of 
the $n_I^{\rm {Baryon}}$ for a given value of the isospin chemical potential $\mu_I$  yields the relation between the isospin chemical potential $\mu_I$ and the isospin number
density of the asymmetric matter $n_I^{\rm {Baryon}}$:
\bee{
n_I^{\rm {Baryon}}=\frac{2}{g_5^2z_m^2}\braa{\mu_I-c_{(3)}}
\ .
\label{nI.vs.mu-hadron_phase}
}
Here in the normal hadron phase the isospin density $n_I$ equals to the density $n_I^{\rm {Baryon}}$ because mesons carrying isospin charge do not condense.

Similarly, for the quark number density we also have
\bee{
n_q=\frac{2}{g_5^2z_m^2}\braa{\mu_q-c_{(0)}}
\ .
}
The baryon number density, $n_B=n_q/N_c$, appears when the baryon chemical potential $\mu_N$ is larger than the mass of nucleon $m_N=939$MeV.
This implies that the $c_{(0)}$ is determined as $c_{(0)}= m_N/N_c $, 
as in Ref.~\cite{Kim:2007zm}.
This argument yields the following relation:
\bee{
n_B=\frac{2}{g_5^2z_m^2N_c^2}\braa{\mu_B-m_N}
\ .
}

Let us next study the $\mu_I$-dependence of the pion mass.
The equations of motion for the pion fluctuation up till the quadratic order in the momentum space are given by 
\bee{
&
-\frac{1}{a \braa{aS^0}^2}\partial_5 \brac{a\braa{aS^0}^2\partial_5 \pi^\pm}
=
E^{\pm}(\mu_I)
 \left[ A_0^\pm +\pi^\pm E^{\pm}(\mu_I)\right]
\ ,
\nn \\ &
\frac{1}{a(aS^0)^2}\partial_5(\frac{a}{g_5^2}\partial_5  A_0^\pm) 
=\left[ A_0^\pm+ \pi^\pm E^{\pm} (\mu_I)\right]
\label{eom for pion}
}
where the fields are parameterized as
\bee{
\pi^{\pm}=\frac{\pi^1\mp i\pi^2}{\sqrt{2}}~,~~A_0^{\pm}=i\frac{A_0^1 \mp i A_0^2}{\sqrt{2}}
}
and 
\bee{
E^{\pm}(\mu_I)=M\pm\mu_I\left(1-\frac{z^2}{z_m^2}\right)\mp c_{(3)} \frac{z^2}{z_m^2}
\label{eigen value}
\ .
}
$S^0$ is the solution of \eqref{S^0} and $M$ is the energy of the static pion.
Equation~(\ref{eom for pion}) together with the boundary conditions,
$\left.\pi\right|_\epsilon=\left.\pd_5\pi\right|_{z_m}=0$, yield
the value of the $M$ as the eigenvalue.
The lowest value of the eigenvalue $M$ is identified with the pion mass, $m_\pi^{*}$.
Here the parameter $c_{(3)}$ is determined as zero by assuming that $\pi^+$ and $\pi^-$ are degenerating at $\mu_I=0$: 
$E^{\pm}(\mu_I)=M\pm\mu_I\braa{1-z^2/z_m^2}$.

\begin{figure}[ht]
 \begin{center}
  \includegraphics[width=70mm]{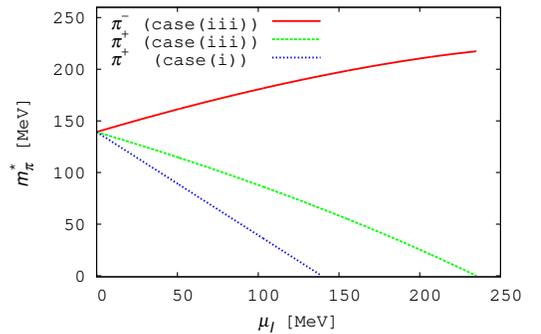}
 \end{center}
 \caption[]{
$\mu_I$ dependence of the pion masses.
The red and green curves show the masses of $\pi^-$ and $\pi^+$, respectively.
We also show the $\mu_I$ dependence of the $\pi^+$ mass in the pure mesonic matter obtained in Ref.~\cite{Nishihara:2014nva} by the blue curve.
}
 \label{fig1}
\end{figure}

Figure~\ref{fig1} shows the $\mu_I$ dependence of the pion mass  in  the normal hadron phase. 
The $\pi^-$ mass drawn by the red curve increases with the isospin chemical potential. 
The $\pi^+$ mass by the green curve, on the other hand, decreases and reaches zero 
at 
$\mu_I=235$MeV, 
 which implies that the $\pi^+$ condenses and that the transition to the pion condensation phase occurs.
We would like to stress that the $\pi^+$ mass here decreases more slowly than the one obtained 
in the pure mesonic matter 
shown by the blue curve.
One can easily see that the critical value of the isospin chemical potential for the phase transition is larger than the pion mass for the pure mesonic case.
This is due to the existence of the baryons in the matter, which can also be understood by an analysis  of the chiral Lagrangian based on the Hidden Local Symmetry as shown in section~\ref{sec:HLS}.

The energy density of the system is defined as $\mc{E}=\Omega+n_q \mu_q+n_I \mu_I $ at zero temperature and
given by
\bee{
\mc{E}=&
                  {} \frac{g_5^2 z_m^2}{4} \braa{n_I}^2
				{}+ \frac{g_5^2 z_m^2N_c^2}{4} \braa{n_B}^2
\ .
}
 Then, the symmetry energy is obtained as
\footnote{Note that this definition of the symmetry energy is different from the one used in Ref.~\cite{Park:2011zp}.}
\bee{
E_{\rm sym}\braa{n_B}\equiv\left.\frac{\pd \braa{\mc{E}/n_B}}{\pd \alpha^2}\right|_{\alpha=0}
=\frac{g_5^2 z_m^2}{16} n_B
}
where $\alpha\equiv\frac{2n_I}{n_B}$.
At the saturation density $n_0=0.16 {\rm fm}^{-3}$, we can estimate $E_{\rm sym}\braa{n_0}=29 {\rm MeV}$ by using \eqref{g_5} and \eqref{parameters}, which is comparable to the empirical value of $32. 3 \pm 1.0$\,MeV \cite{Zhang:2013wna}.
In  Refs.~\cite{Chen:2004si,Shetty:2007zg}, the value of the parameter $\gamma$ defined as $E_{\rm sym}(n_B)=E_{\rm sym}(\rho_0)\braa{\frac{n_B}{n_0}}^\gamma$ is estimated as $\gamma=0.55$ - $0.69$, which is different from the result of the present analysis, $\gamma=1$.
The slope parameter of the symmetry energy is give by
\bee{
L\equiv 
3n_0\left.\frac{\pd E_{\rm sym} (n_B)}{\pd n_B}\right|_{n_B=n_0}
=3n_0\frac{g_5^2 z_m^2}{16}
}
and its value is estimated as $L=87$MeV, where its empirical value is known as $45.2 \pm 10.0$\,MeV \cite{Zhang:2013wna}.
We may understand that the deviations of values of $\gamma$ and $L$ are caused by  the next leading order in the large $N_c$ expansion.

In Case-(ii), as we stated
in section~\ref{sec:model}, $2n_I^{\rm Baryon}$ and $n_B$ are equal to $n_p$, which
 leads to
the following solutions of \eqref{eom for phi} in the normal hadron phase, $b=\theta=0$:
\bee{
\varphi^3=&-\frac{g_5^2 n_I^{\rm {Baryon}}}{2} z^2=-2\frac{g_5^2 n_p}{2} z^2
\ ,\nn\\
\varphi^0=&-\frac{g_5^2 n_q}{2} z^2=-N_c\frac{g_5^2 n_p}{2} z^2
\ .
}
Now, 
the grand potential density is given by
\bee{
\Omega 
\supset&
                  {} -\int_\epsilon^{z_m} dz~\frac{a}{2g_5^2}\brac{\left(\partial_5 V_0^0 \right)^2 +\left(\partial_5 V_0^3 \right)^2 }
\nn\\
&-n_I^{\rm {Baryon}} \left.V_0^3\right|_{z_m-\delta z}
-n_q \left.V_0^0\right|_{z_m-\delta z}
\nn\\
=&
				{}\frac{g_5^2}{4}\braa{{N_c^2}+\frac{1}{4}}z_m^2{n_p^2} 
-\braa{ \mu_p-N_c c_{(0)}-\frac{c_{(3)}}{2}}n_p
\ ,
}
where we used $2n_I^{\rm {Baryon}}=n_B=n_p$ and $\mu_p= \mu_B+\frac{\mu_I}{2}$.
Minimizing this in terms of $n_p$, 
we have
\bee{
n_p=\frac{2}{g_5^2z_m^2}\frac{4}{1+4N_c^2}\braa{\mu_p-c_{(0)}-\frac{c_{(3)}}{2}}
}
Thus, we
set $c_{(0)}+\frac{c_{(3)}}{2}=m_N$ 
because the proton density $n_p$ must vanish as $\mu_p \ra m_N$.

In Case-(ii), the pion mass depends on not only $\mu_I$ but also $\mu_B$ through 
\bee{
E^{\pm}(\mu_I)=&M\pm\mu_I\braa{1-\frac{1}{1+4N_c^2}\frac{z^2}{ z_m^2}
}
\nn\\&
\mp2\frac{1}{1+4N_c^2}\braa{\mu_B-m_N}
\frac{z^2}{ z_m^2}
\mp c_{(3)}\frac{z^2}{ z_m^2}
}
in \eqref{eigen value}.
In the limit $\mu_I\ra 0$ at $\mu_B=m_N$
\footnote{
We can not take the limit $\mu_I\ra 0$ at $\mu_B\neq m_N$ in Case-(ii).
} 
the degeneration of the charged pions gives us $c_{(3)}=0$.
The $\mu_I$ dependence of the pion mass is the almost same 
as that in 
the pure mesonic case (Case-(i)) because 
the difference is suppressed by 
the factor $\frac{1}{1+4N_c^2}=\frac{1}{37}$.

\section{Pion Condensation Phase}
\label{sec:pion condensation}

Next, we study the equation of state 
 in the asymmetric nuclear matter.
First, we will perform the following analysis in Case-(ii) and in Case-(iii),
separately, and show the results of in Case-(iii).
In the last of this section, our results on the ($\mu_I$, $\mu_B$) plane will be shown, which are given by combining   the result of each Case.

\newpage
\begin{widetext}
{}From the Lagrangian
\eqref{Lagrangian}, the equations of motion are obtained as
\bee{
\partial_5 \left( -a^3 \partial_5 S^0 \right) + a^3 S^0 \left(\partial_5 b\right)^2
 -3a^5S^0
-a^3S^0\left[\sin^2 b \; ( \varphi^3 +\mu_I )^2
+\theta\sin 2b \;\sin \zeta \; (\varphi^3 +\mu_I) 
+\theta^2 - \theta^2 \sin^2 b \; \sin^2 \zeta \right] =&0
\ ,
\nn \\ 
\partial_5 \left( - a^3 \left(S^0\right)^2 \partial_5 b \right) 
-\frac{a^3\left(S^0\right)^2}{2}\left[\sin 2b \; \left\{( \varphi^3 +\mu_I )^2
- \theta^2 \sin^2 \zeta\right\}+2\theta \cos 2b \;\sin \zeta 
~( \varphi^3 +\mu_I )\right]=&0
\ ,
\nn \\ 
\partial_5\left(\frac{a}{g_5^2}\partial_5 \theta \right)
-\frac{a}{g_5^2}\theta \left( \partial_5 \zeta \right)^2
-\frac{a^3\left(S^0\right)^2}{2}\left[\sin 2b \; \sin \zeta 
~( \varphi^3 +\mu_I )+ 2\theta \left\{ 1 -  \sin^2 b \;\sin^2 \zeta \right\}\right]=&0
\ ,
\nn \\ 
\partial_5\left(\frac{a}{g_5^2}\theta^2\partial_5 \zeta \right)
-\frac{a^3\left(S^0\right)^2}{2}\left[\theta \sin 2b \;\cos \zeta \;\;( \varphi^3 +\mu_I )
- \theta^2 \sin^2 b\sin 2 \zeta\right]=&0
\ ,
\nn \\ 
\partial_5 \left( \frac{a}{g^2_5} \partial_5 \varphi^3 \right)
-\frac{a^3\left(S^0\right)^2}{2}\left[2\sin^2 b 
\;( \varphi^3 +\mu_I )+\theta\sin 2b \;\sin \zeta \right]=&0
\ .
\label{EOM}
}
\end{widetext}
These differential equations are solved with the boundary conditions listed in Table \ref{Boundary conditions}.
\begin{table}[h]
 \begin{center}
 \begin{tabular}{ccc}\hline\hline
Variables	&	UV	&	IR	
\\\hline
$S^0$	&	$\frac{S^0}{z} |_\epsilon=m_q$	&	
$ \partial_5 S^0 |_{z_m}=
\left. -\frac{S^0}{2z_m} \left(\lambda \left(S^0\right)^2 - 2 m^2 \right) \right\vert_{z_m}$
\\
$b$	&	$b |_\epsilon=0$	&	$\partial_5 b |_{z_m}=0$
\\
$\theta$	&	$\theta |_\epsilon=0$	&	$\partial_5 \theta |_{z_m}=0$
\\
$\zeta$	&	$\zeta |_\epsilon=\frac{\pi}{2}$	&	$\partial_5 \zeta |_{z_m}=0$
\\
$\varphi^3$	&	$\varphi^3 |_\epsilon=0$	&	
\begin{tabular}{ll}
$\varphi^3 |_{z_m}=-\mu_I$ & in Case-(iii) \\
(IR condition) & in Case-(ii) \\
\end{tabular}
\\
\hline\hline
 \end{tabular}
 \end{center}
\caption{Boundary conditions for the relevant wave functions.
(IR condition) implies that $N_c^2\left.\pd_5 \varphi^3\right|_{z_m} +\frac{1}{2}\left.\varphi^3\right|_{z_m}+\mu_B+\frac{\mu_I}{2}-m_N=0$ is satisfied.
}
\label{Boundary conditions}
\end{table}
We note that, in Case-(iii), the IR boundary condition for $\varphi^3$ and the value of $n_I^{\rm Baryon}$ are determined by 
the minimization condition for the grand potential 
\bee{
0 =
\frac{\pd \Omega}{\pd n_I^{\rm {Baryon}}}=&-\int dzV_0^3 
\delta\left(z-z_m+\delta z\right)
\nn\\
=& -\mu_I-\left.\varphi^3 \right|_{z_m-\delta z}
\ ,
}
and the condition in \eqref{BD for V}
\bee{
n_I^{\rm {Baryon}}=\left.-\frac{1}{g_5^2}\frac{\partial_5 \varphi^3 }{z}\right|_{z_m}
\ .
}
In Case-(ii), 
on the other hand,
the condition
 $\pd \Omega / \pd n_p=0$ together with 
the solution of $\varphi^0$ and the condition in \eqref{BD for V}
leads to
\begin{equation}
N_c^2\left.\pd_5 \varphi^3\right|_{z_m} +\frac{1}{2}\left.\varphi^3\right|_{z_m}+\mu_B+\frac{\mu_I}{2}-m_N=0
\ .
\end{equation}
It should be noticed that the condition provides the $\mu_B$ dependence of the isospin number density $n_I$ in Case-(ii), although the coupled equations of motion in Eq.~(\ref{EOM}) do not include $\mu_B$.
On the other hand, in Case-(iii), neither the boundary conditions nor  the equations of motion have $\mu_B$ dependence, which implies that the isospin number density is independent of the baryon number chemical potential.

Now, let us study the isospin number density in Case-(iii), which
is defined by
\bee{
n_I=-\frac{\pd \Omega}{\pd \mu_I}=n_I^{\rm {Meson}}+n_I^{\rm {Baryon}}
}
where the $n_I^{\rm {Meson}}$ expresses the mesonic contribution to the isospin number density given by
\bee{
n_I^{\rm {Meson}}=\int dz \frac{a^3\left(S^0\right)^2}{2}\left[2\sin^2 b ( \varphi +\mu_I )+\theta\sin 2b \sin \zeta \right]
\ .
}

\begin{figure}[ht]
 \begin{center}
   \includegraphics[width=70mm]{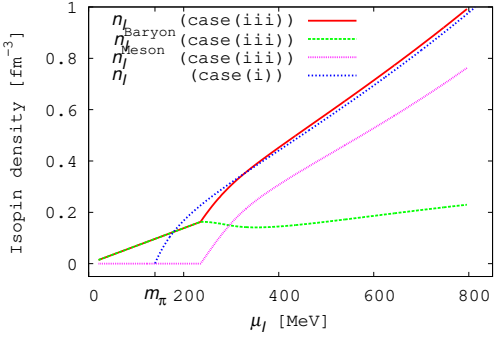}
 \end{center}
 \caption[]{
Equation of state 
between the isospin chemical potential $\mu_I$ and the isospin number density $n_I$ drawn by the red curve.
The blue and green curves show the $\mu_I$ dependences of the mesonic contribution $n_I^{\rm {Meson}}$ and the baryonic contribution $n_I^{\rm {Baryon}}$, respectively.
The pink
dots show the result shown in Ref.~\cite{Nishihara:2014nva} for pure mesonic matter.}
 \label{fig2}
\end{figure}

In Fig.~\ref{fig2}, we show the resultant equation of state between the isospin density and the isospin chemical potential  obtained by solving \eqref{EOM}.
For $\mu_I < 235$\,MeV there is no pion condensation, so that the isospin number density increases linearly with the chemical potential following \eqref{nI.vs.mu-hadron_phase} as drawn by the red curve in Fig.~\ref{fig2}. At $\mu_I^c = 235$MeV the phase transition occurs from the normal hadron phase to the pion condensation phase.
This critical chemical potential $\mu_I^c = 235$\,MeV is consistent with the one determined from the pion mass shown in \figref{fig1}, but the value
is larger than the critical value for the pure mesonic matter, for which $\mu_I^c = m_\pi$~\cite{Nishihara:2014nva} as seen by the green curve in Fig.~\ref{fig2}.
This delay of the phase transition is due to the existence of the baryons, which can also be understood by an analysis of the four dimensional chiral Lagrangian as shown in the next section.

In the pion condensation phase, the pion contribution to the isospin number density increases monotonically with the chemical potential as shown by the pink curve in Fig.~\ref{fig2}, while the baryonic contribution by the blue curve is almost constant: $n_I^{\rm {Baryon}} \sim  0.2\ {\rm fm}^{-3}$.
As a result the mesonic contribution dominates the isospin number density.
This implies that the energy provided by the isospin chemical potential is mostly used for generating the pion condensation rather than converting the neutron into proton.

\begin{figure}[ht]
 \begin{center}
   \includegraphics[width=70mm]{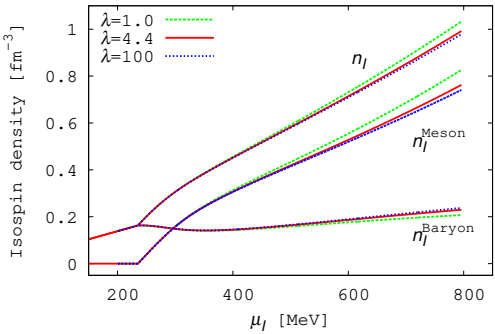}
 \end{center}
 \caption[]{
Dependence of the equation of state on the value of $\lambda$. 
The green, red and blue curves are for $\lambda=1.0, 4.4$ and $100$, respectively.
}
 \label{fig7}
\end{figure}

Figure \ref{fig7} shows the dependence of the equation of state on the scalar meson mass.
The value of parameter $\lambda$ is determined from the mass of the $a_0$ meson,
where $\lambda=1.0$, $4.4$ and $100$ correspond to the $m_{a_0}=610, 980$ and $1210$ MeV, respectively.
We find that the critical value of the isospin chemical potential is independent of $\lambda$ and the behavior of the equation of state is not sensitive to the value of $\lambda$.

As we stated in the introduction, the existence of the isospin chemical potential $\mu_I$ explicitly breaks the chiral symmetry group SU$(2)_R\times$ SU$(2)_L$  down to its
subgroup $\mbox{U}(1)_R^{(3)} \times \mbox{U}(1)_L^{(3)} = \mbox{U}(1)_V^{(3)} \times \mbox{U}(1)_A^{(3)}$, where the superscript $^{(3)}$ implies that the generator $T_3$ of SU(2) is used for the U(1) as $\exp[ i \theta_V T_3 ] \in \mbox{U}(1)_V^{(3)}$.
For studying the order parameters for the phase transition, we define the following $\pi$-condensate and the ``$\sigma$''-condensate~\cite{Nishihara:2014nva}:
\begin{eqnarray}
\langle\pi^a\rangle 
&\equiv&
\frac{1}{2}\mathrm{Tr} \left[i\sigma^a a\left(\partial_5 \frac{X}{z}\right)
+{\rm h.c.}\right]_\epsilon 
=\langle \bar{q} \gamma_5 \sigma^a q \rangle
\ ,
\nonumber \\
\langle\sigma\rangle
&\equiv& 
\frac{1}{2}\mathrm{Tr} \left[ a\left(\partial_5 \frac{X}{z}\right)
+{\rm h.c.}\right]_\epsilon 
= \langle \bar{q} q \rangle 
\ .
\end{eqnarray}

\begin{figure}[ht]
 \begin{center}
   \includegraphics[width=70mm]{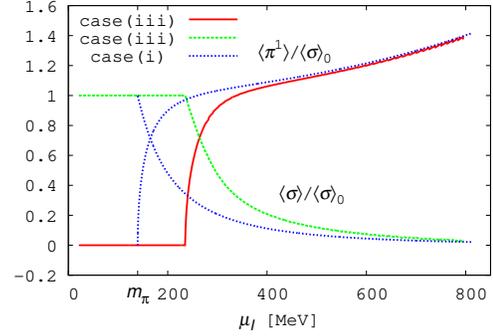}
 \end{center}
 \caption{
Dependence of the ``$\sigma$''-condensate $\langle \sigma \rangle$ and the $\pi$-condensate $\langle \pi \rangle$ on the isospin chemical potential $\mu_I$.
The condensates are scaled by the ``$\sigma$''-condensate at the vacuum indicated by $\langle \sigma \rangle_0$.
The blue curves are the dependence of $\langle \sigma \rangle$ and $\langle \pi \rangle$ shown in Ref.~\cite{Nishihara:2014nva} for the pure mesonic matter.
}
 \label{fig3}
\end{figure}

We plot the ``$\sigma$''-condensate and the $\pi$-condensate obtained by the present analysis in Fig.~\ref{fig3}, together with those condensates for the pure mesonic matter. 
This figure shows that the present behavior is quite similar to the  previous one except the difference of the phase transition point:
In the normal hadron phase the ``$\sigma$''-condensate exists, which leads to the break down of 
the $\mbox{U}(1)_A^{(3)}$  symmetry, but $\pi$-condensate is zero.
At the phase transition point, the $\pi$-condensate appears, which spontaneously breaks the $\mbox{U}(1)_V^{(3)}$ symmetry, while the ``$\sigma$''-condensate starts to decrease very rapidly.
For large $\mu_I$, the ``$\sigma$''-condensate is almost zero while the $\pi$-condensate keeps increasing.

\begin{figure}[ht]
 \begin{center}
   \includegraphics[width=70mm]{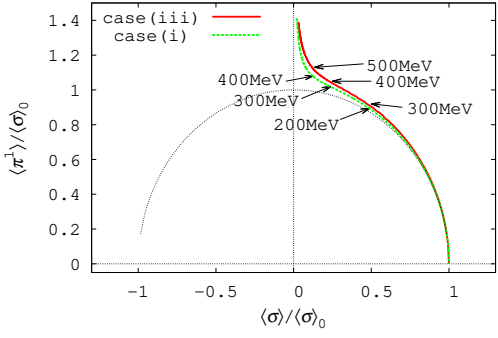}
 \end{center}
 \caption{
Chiral circle shown by red curve.
The black curve is an unit circle and the green curve is the chiral circle for the pure mesonic matter.
}
 \label{fig4}
\end{figure}

We next show the chiral circle in Fig.~\ref{fig4}.
The red solid curve shows that the behavior for the nuclear matter is quite similar to the one for the pure mesonic matter shown by the green dotted line:
Although the ``$\sigma$''-condensate decreases and the $\pi$-condensate increases, the chiral condensate defined by
\begin{equation}
\tilde{\sigma} \equiv \sqrt{ \langle \sigma \rangle^2 + \langle \pi^a \rangle^2 }
\end{equation}
stays constant until about 150\,MeV above the critical chemical potential.
In the large $\mu_I$ region, the chiral condensate $\tilde{\sigma}$ grows very rapidly.
This implies that the enhancement of the chiral symmetry breaking occurs in the asymmetric nuclear matter, similarly to the one in the pure mesonic matter as shown in Ref.~\cite{Nishihara:2014nva}.

\begin{figure}[ht]
 \begin{center}
   \includegraphics[width=70mm]{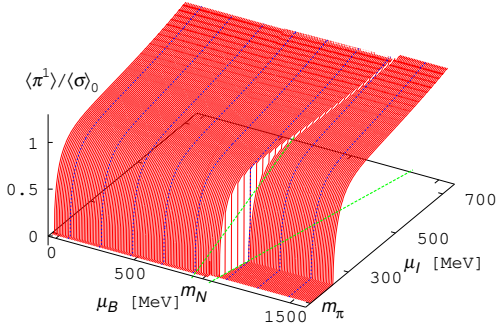}
 \end{center}
\vspace{0.2cm}
 \caption{
 $\frac{\langle\pi^1\rangle}{\langle\sigma\rangle_0}$ vs. $\mu_B$ vs. $\mu_I$.
The green lines are the boundaries between Case-(i) and Case-(ii) and between Case-(ii) and Case-(iii).
}
 \label{fig8}
\end{figure}

\begin{figure}[ht]
 \begin{center}
   \includegraphics[width=70mm]{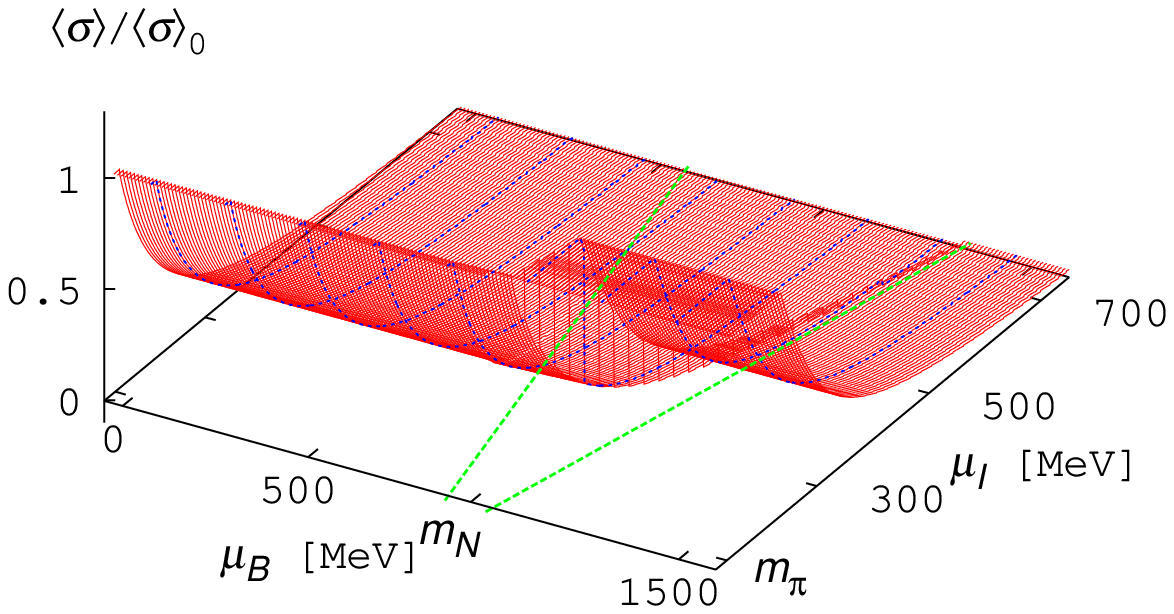}
\vspace{0.2cm}
 \end{center}
 \caption{
 $\frac{\langle\sigma\rangle}{\langle\sigma\rangle_0}$ vs. $\mu_B$ vs. $\mu_I$.
The green lines are the boundaries between Case-(i) and Case-(ii) and between Case-(ii) and Case-(iii).
}
 \label{fig9}
\end{figure}

\begin{figure}[ht]
 \begin{center}
   \includegraphics[width=70mm]{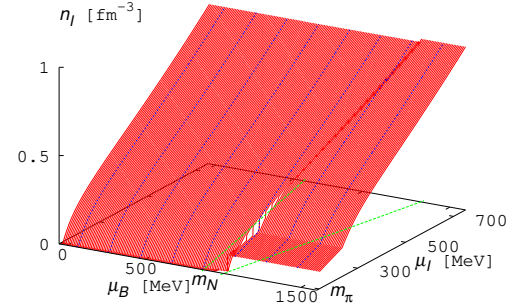}
 \end{center}
 \caption{
 $n_I$ vs. $\mu_B$ vs. $\mu_I$.
The green lines are the boundaries between Case-(i) and Case-(ii) and between Case-(ii) and Case-(iii).
}
 \label{fig10}
\end{figure}

Next, we 
study the equation of state and the condensates in Case-(ii) as well as those in Case-(i) in a similar way.
The resultant
$\langle\pi^1\rangle$, $\langle\sigma\rangle$ and $n_I$ 
in the entire ($\mu_B$, $\mu_I$) plane
are shown in Fig.~\ref{fig8},~\ref{fig9} and \ref{fig10}, respectively.
The green lines in these figures show the boundary between Case-(i) and Case-(ii) and that between Case-(ii) and Case-(iii), which are corresponding to $\mu_p=m_N$ and $\mu_n=m_N$. 
Figure~\ref{fig8} shows that there is the first order transition on the boundary between the pion condensation phase in Case-(ii) and the normal hadron phase in Case-(iii).  
In Fig.~\ref{fig9}, we see that $\langle\sigma\rangle$ decreases discontinuously at  the first order transition line in response to sudden increase of $\langle\pi^1\rangle$ in Fig.~\ref{fig8}.
Figure~\ref{fig10} shows the equation of state on the ($\mu_B$, $\mu_I$) plane. 
In these figures, the values of $\langle\pi^1\rangle$, $\langle\sigma\rangle$ and $n_I$ drastically change on $\mu_n=m_N$ which is the boundary between Case-(ii) and Case-(iii).

\section{An analysis by the chiral Lagrangian based on the Hidden Local Symmetry}
\label{sec:HLS}

In this section, we show that the delay of the phase transition to the pion condensation phase is understood as the baryonic matter effect in the framework of the four dimensional chiral Lagrangian including the $\rho$ meson based on the hidden local symmetry (HLS)~\cite{Bando:1987br,Harada:2003jx}.

The mesonic part of the HLS Lagrangian is given by 
\bee{
\mathcal{L}=&F_\pi^2 \tr\brac{\hat{\alpha}_{\perp \mu}\hat{\alpha}_\perp^\mu}
+aF_\pi^2 \tr\brac{\hat{\alpha}_{\parallel \mu}\hat{\alpha}_\parallel^\mu}
\nn\\
&+\frac{F_\pi^2}{4}\tr\brac{\xi_L \chi \xi_R^\dagger+\xi_R \chi^\dagger \xi_L^\dagger}
-\frac{1}{2g^2}\tr\brac{V_{\mu\nu}V^{\mu\nu}}
}
where $\chi$ is an external field which has the expectation value corresponding to the pion mass, $\brae{\chi}=m_\pi^2 {\bf 1}$.
The $\hat{\alpha}_{\perp \mu}$ and $\hat{\alpha}_{\parallel \mu}$ are defined as
\bee{
\hat{\alpha}_{\perp, \parallel \mu}=&\frac{D_\mu\xi_{L} \cdot \xi_{L}^\dagger\pm D_\mu\xi_{R} \cdot\xi_{R}^\dagger}{2i}
}
where $\xi_{L,R}$ are the fields including pions, $V_\mu$ is the gauge field including the rho and omega mesons and the covariant derivative of these fields are
\bee{
D_\mu\xi_{L}=&\partial_\mu \xi_{L}-iV_\mu\xi_{L}+i\xi_{L}\mathcal{L}_\mu
\ ,\nn\\
D_\mu\xi_{R}=&\partial_\mu \xi_{R}-iV_\mu\xi_{R}+i\xi_{R}\mathcal{R}_\mu
\ .
}
The baryon and isospin chemical potentials, $\mu_B$ and $\mu_I$, are introduced as the expectation value of the time component of the external gauge fields: $\brae{\mc{L}_0}=\brae{\mc{R}_0}=\frac{\mu_B}{2}\sigma^0+\frac{\mu_I}{2}\sigma^3$.

Here we introduce the following terms including the baryons explicitly:
\bee{
\bar{N} i \gamma^\mu D_\mu N+G\bar{N}\gamma^\mu \hat{\alpha}_{\mu \parallel}N
}
where $N$ is the baryon field and $D_\mu$ is a covariant derivative defined as $D_\mu N = \left( \partial_\mu - i V_\mu \right) N$.
We replace the bilinear baryon fields by the mean field as
\begin{equation}
\braa{V_0^{3} +G\hat{\alpha}_{\parallel 0}^3}n_I^{\rm {Baryon}}
+\braa{V_0^{0} +G\hat{\alpha}_{\parallel 0}^0}n_B 
\end{equation}
where $\hat{\alpha}_{\parallel 0}^{(0,3)}=\tr\braa{\hat{\alpha}_{\parallel 0}\sigma^{(0,3)} }$, $V_0^{3}$ is the time component of the neutral rho meson and $V_0^{0}$ is the time component of the omega meson.

\begin{figure}[ht]
 \begin{center}
  \includegraphics[width=70mm]{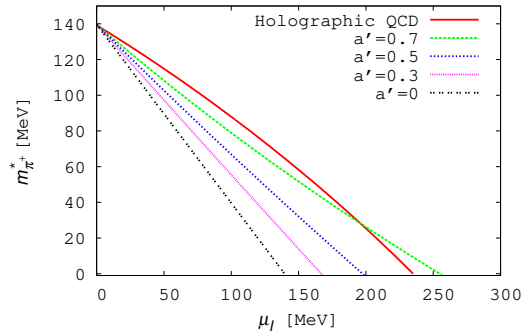}
 \end{center}
 \caption{
The $\mu_I$ dependence of the $\pi^+$ mass obtained from the chiral Lagrangian.
The green, blue and pink curves are the results for $a'=0.7, 0.5$ and $0.3$ respectively. The $\pi^+$ mass given from the analysis of the holographic QCD model is indicated by the red curve. 
We also show the $\mu_I$ dependence of the $\pi^+$ mass in the pure mesonic matter obtained in Ref.~\cite{Nishihara:2014nva} by the dotted black line.
}
 \label{fig6}
\end{figure}

Taking the unitary gauge of the HLS and integrating out the rho and omega mesons and assuming the rotational symmetry, we obtain the following effective Lagrangian for the pion coupling to the baryonic sources:
\bee{
\mathcal{L}=&F_\pi^2 \tr\brac{\hat{\alpha}_{\perp 0}\hat{\alpha}_\perp^0}
+\frac{F_\pi^2}{4}\tr\brac{\xi_L \chi \xi_R^\dagger+\xi_R \chi^\dagger \xi_L^\dagger}
\nn\\&
-\frac{1}{2a'F_\pi^2}n_B^2-\frac{1}{2a'F_\pi^2}\braa{n_I^{\rm {Baryon}}}^2
+\mu_Bn_B
+{\alpha}_{\parallel 0}^3n_I^{\rm {Baryon}}
\label{the effective Lagrangian based on the HLS}
}
where $a'\equiv\frac{a}{\braa{1-G}^2} $ \footnote{Since the value of the parameter $a$ is known as about two in the HLS~\cite{Harada:2003jx}, $a'=\frac{a}{\braa{1-G}^2} $ is larger than zero.} and ${\alpha}_\parallel^\mu=\hat{\alpha}_\parallel^\mu+V^\mu$.
Existence of the terms in the last line of \eqref{the effective Lagrangian based on the HLS} causes the deviation from the result obtained by the $\mc{O}(p^2)$ chiral Lagrangian without the baryonic sources, which delays the transition to the pion condensation comparing to of the pure mesonic analysis.
Figure \ref{fig6} shows the relation between the pion mass and the isospin chemical potential for $a'=0.7$ (green), $0.5$ (blue), $0.3$ (pink) and of the holographic QCD model (the red curve). The dotted black line corresponds to the case for the pure pion matter, $a'=0$.
This figure shows that the point at which the curve reaches zero depends on the value of $a'$.
The critical value of the isospin chemical potential for $0 < a' < 1$
 is larger than the pion mass, which implies that
delay of the transition is understood by using a model based on the HLS with the baryonic  sources.

\section{A summary and discussions}
\label{sec:summary}

We introduced a baryonic source at the IR boundary coupling to the iso-triplet vector meson in the hard wall holographic QCD mode, and studied the pion condensation in the asymmetric nuclear matter.
We showed that the phase transition from the normal matter to the pion condensation phase is delayed in the asymmetric nuclear matter compared with the pure mesonic matter.
Furthermore, our result shows that the meson contribution to the isospin number density increases with the chemical potential, while the baryon contribution stays constant. 
We would like to stress
that the chiral symmetry breaking is enhanced in the asymmetric nuclear matter as in the pure mesonic matter.

\begin{figure}[ht]
 \begin{center}
   \includegraphics[width=60mm]{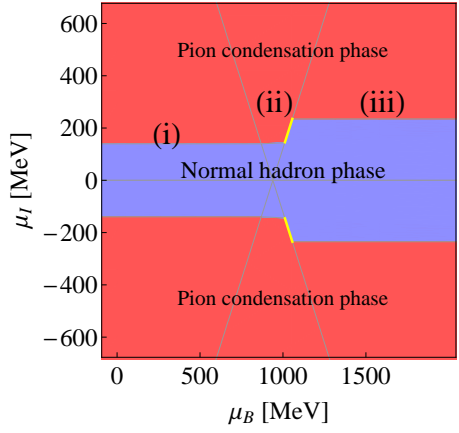}
 \end{center}
 \caption{
Phase diagram: $\mu_B$ vs. $\mu_I$.
The blue and red area express the hadron phase and the pion condensation phase respectively.
}
 \label{fig5}
\end{figure}

We show the phase diagram obtained from the present analysis in \figref{fig5}, where the blue and red area express the hadron phase and the pion condensation phase respectively.
The phase transition is of the second order except on the yellow line expressing the first order.
In Case-(i), the phase transition to the pion condensation occurs at which the isospin chemical potential is equal to the pion mass as shown in Ref.~\cite{Nishihara:2014nva}.
On the other hand, in Case-(iii), done by the present analysis, the critical point of the transition is delayed compared 
with in Case-(i).
A similar delay also occurs in Case-(ii), although the effect is very tiny and it is hard to see in Fig.~\ref{fig5}.

The model which we used in section \ref{sec:HLS} 
explicitly includes 
the rho and omega mesons.
The existence of the rho meson is essential for the delay of the phase transition.
This indicate that the phase transition point in the NJL model may be changed 
by including
the following vector 4-Fermi interaction~\cite{Asakawa:1989bq,Kitazawa:2002bc,Fukushima:2008is,Zhang:2009mk,Bratovic:2012qs,Zhang:2013oia}:
\bee{
g_v\brac{\braa{\bar{\psi}\sigma^a\gamma_\mu \psi }^2+\braa{\bar{\psi}\sigma^a\gamma_\mu \gamma_5\psi }^2}
}
where $g_v$ is a positive coupling constant, $\psi$ is a quark field and $\sigma^a$ are Pauli matrices in the flavor space.

In the present analysis, we put the baryonic charge at the IR boundary.  In more general case, the charge is spread into the bulk by the gauge interaction.  Furthermore, the coupling of the baryon to the scalar mesons is not included.
Such effects could be included by the holographic mean field approach \cite{Harada:2011aa,He:2013gta}, which is left for future publication.

References \cite{Lee:2013oya,Park:2011zp,Jo:2009xr} studied
the asymmetric matter in the hard wall holographic QCD model.
Our results for the meson mass splitting and the symmetry energy are comparable to their results.

\section*{ACKNOWLEDGEMENTS}
The authors would like to thank Kenji Fukushima for stimulating discussion on the symmetry structure.
This work was supported in part by Grant-in-Aid for Scientific Research
on Innovative Areas (No. 2104) ``Quest on New Hadrons with Variety of Flavors'' from MEXT, and 
the JSPS Grant-in-Aid for Scientific Research
(S) No.~22224003, (c) No.~24540266.

\appendix

\section{Equations of motion}
\label{app:EOM}

At the vacuum, non zero value of $S^0$ brakes the chiral symmetry to the vector part of its symmetry.
The iso-singlet scalar field $S^0$ satisfies the following equation of motion (EOM) and the boundary conditions:
\bee{
\pd_5 a^3 \pd_5 S^0+3 a^5S^0=&0
\ ,
\nn\\
m_q=
\left.\frac{S^0}{z}\right|_{\epsilon}
\ ,
\nn\\
\brac{  \pd_5 S^0 +\frac{S^0}{2z_m} \left(\lambda \left(S^0\right)^2 - 2 m^2 \right) }_{z_m}=0
\label{S^0}
}
where the $m_q$ corresponds to the explicit braking of the chiral symmetry due to the current quark mass.

\begin{widetext}
Using the assumptions given in section~\ref{sec:model} and the variables parameterized in Eqs. (\ref{parameterize4}) and (\ref{parameterize5}), the Lagrangian $\mc{L}$ is written as
\bee{
\mc{L}=&\mc{L}_{1}+\mc{L}_{2}
\ ,
\nn\\
\mc{L}_{1}
=&\frac{a^3}{2}\left[-\left(\partial_5 S^0\right) ^2
-\left(S^0\right)^2\left(\partial_5 b\right)^2
\right]
                 {} +\frac{3a^5}{2} \left(S^0\right)^2
\nn \\ &
{} +\frac{a^3\left(S^0\right)^2}{2}\left[\sin^2 b ~( V_0^3 )^2
			+\theta\sin 2b \sin \zeta ~V_0^3 
		+\theta^2- \theta^2 \sin^2 b \sin^2 \zeta\right]
\nn \\ &
                  {} +\frac{a}{2g_5^2}\left[\left(\partial_5 V_0^3 \right)^2
	+ \left(\partial_5 \theta \right)^2
		+\theta^2 \left(\partial_5 \zeta \right)^2 \right]
				+\rho_I V_0^3\delta (z-z_m+\delta z)
\ ,
\nn\\
\mc{L}_{2}=&
                {} \frac{a}{2g_5^2}\left(\partial_5 V_0^0 \right)^2 
					{}+\rho_q V_0^0\delta (z-z_m+\delta z)
\label{Lagrangian}
}
where
\bee{
e^{i\pi^a \sigma^a }=&\cos b +i \sin b \sigma^1
\ ,
\nn\\
A_0^a=&\braa{\theta \cos \zeta,\theta \sin \zeta,0}
\ .
}
\end{widetext}
For convenience, we fixed $\pi^2=0$ by using the isospin symmetry U$(1)_I$ which is the subgroup of U$(1)^{3}_L\times$ U$(1)^3_R$.


\begin{thebibliography}{100}

\bibitem{Lattimer:2006xb} 
See, e.g. 
	  J.~M.~Lattimer and M.~Prakash,
	  Phys.\ Rept.\  {\bf 442}, 109 (2007)
and references therein.

\bibitem{Demorest:2010bx} 
  P.~Demorest, T.~Pennucci, S.~Ransom, M.~Roberts and J.~Hessels,
  Nature {\bf 467}, 1081 (2010).

\bibitem{Antoniadis:2013pzd} 
  J.~Antoniadis, P.~C.~C.~Freire, N.~Wex, T.~M.~Tauris, R.~S.~Lynch, M.~H.~van Kerkwijk, M.~Kramer and C.~Bassa {\it et al.},
  Science {\bf 340}, 6131 (2013).

\bibitem{Fukushima:2010bq} 
  K.~Fukushima and T.~Hatsuda,
  Rept.\ Prog.\ Phys.\  {\bf 74}, 014001 (2011).

\bibitem{Fukushima:2013rx} 
  K.~Fukushima and C.~Sasaki,
  Prog.\ Part.\ Nucl.\ Phys.\  {\bf 72}, 99 (2013).

 \bibitem{Sasaki:2010jz} 
  T.~Sasaki, Y.~Sakai, H.~Kouno and M.~Yahiro,
  Phys.\ Rev.\ D {\bf 82}, 116004 (2010).

\bibitem{Zhang:2006gu} 
  Z.~Zhang and Y.~X.~Liu,
  Phys.\ Rev.\ C {\bf 75}, 064910 (2007).

\bibitem{Toublan:2003tt} 
  D.~Toublan and J.~B.~Kogut,
  Phys.\ Lett.\ B {\bf 564}, 212 (2003).

\bibitem{Barducci:2004tt} 
  A.~Barducci, R.~Casalbuoni, G.~Pettini and L.~Ravagli,
  Phys.\ Rev.\ D {\bf 69}, 096004 (2004).

\bibitem{He:2005nk} 
  L.~y.~He, M.~Jin and P.~f.~Zhuang,
  Phys.\ Rev.\ D {\bf 71}, 116001 (2005).

\bibitem{Mu:2010zz} 
  C.~f.~Mu, L.~y.~He and Y.~x.~Liu,
  Phys.\ Rev.\ D {\bf 82}, 056006 (2010).

\bibitem{He:2006tn} 
  L.~He, M.~Jin and P.~Zhuang,
  Phys.\ Rev.\ D {\bf 74}, 036005 (2006).

\bibitem{He:2005tf} 
  L.~y.~He, M.~Jin and P.~f.~Zhuang,
  Mod.\ Phys.\ Lett.\ A {\bf 22}, 637 (2007).

\bibitem{Zhang:2013oia} 
  Z.~Zhang and H.~P.~Su,
  Phys.\ Rev.\ D {\bf 89}, 054020 (2014).

\bibitem{Andersen:2007qv} 
  J.~O.~Andersen and L.~Kyllingstad,
  J.\ Phys.\ G {\bf 37}, 015003 (2009).

\bibitem{Lee:2013oya}
 B.~-H.~Lee, S.~Mamedov, S.~Nam and C.~Park,
 JHEP {\bf 1308}, 045 (2013).

\bibitem{Parnachev:2007bc} 
  A.~Parnachev,
  JHEP {\bf 0802}, 062 (2008).

\bibitem{Klein:2003fy} 
  B.~Klein, D.~Toublan and J.~J.~M.~Verbaarschot,
  Phys.\ Rev.\ D {\bf 68}, 014009 (2003).

\bibitem{Nishida:2003fb} 
  Y.~Nishida,
  Phys.\ Rev.\ D {\bf 69}, 094501 (2004).

\bibitem{Toublan:2004ks} 
  D.~Toublan and J.~B.~Kogut,
  Phys.\ Lett.\ B {\bf 605}, 129 (2005).

\bibitem{Abuki:2013vwa} 
  H.~Abuki,
  Phys.\ Rev.\ D {\bf 87}, 094006 (2013).

\bibitem{Nishihara:2014nva} 
  H.~Nishihara and M.~Harada,
  Phys.\ Rev.\ D {\bf 89}, 076001 (2014).

\bibitem{Son:2000xc} 
  D.~T.~Son and M.~A.~Stephanov,
  Phys.\ Rev.\ Lett.\  {\bf 86}, 592 (2001).

\bibitem{Kim:2007zm} 
  K.~-Y.~Kim, S.~-J.~Sin and I.~Zahed,
  JHEP {\bf 0801}, 002 (2008).

\bibitem{Bando:1987br} 
  M.~Bando, T.~Kugo and K.~Yamawaki,
  Phys.\ Rept.\  {\bf 164}, 217 (1988).

\bibitem{Harada:2003jx} 
  M.~Harada and K.~Yamawaki,
  Phys.\ Rept.\  {\bf 381}, 1 (2003).

\bibitem{Erlich:2005qh} 
  J.~Erlich, E.~Katz, D.~T.~Son and M.~A.~Stephanov,
  Phys.\ Rev.\ Lett.\  {\bf 95}, 261602 (2005).

\bibitem{Da Rold:2005zs} 
  L.~Da Rold and A.~Pomarol,
  Nucl.\ Phys.\ B {\bf 721}, 79 (2005).

\bibitem{DaRold:2005vr} 
  L.~Da Rold and A.~Pomarol,
  JHEP {\bf 0601}, 157 (2006).

\bibitem{Park:2011zp} 
  C.~Park,
  Phys.\ Lett.\ B {\bf 708}, 324 (2012).

\bibitem{Zhang:2013wna} 
  Z.~Zhang and L.~-W.~Chen,
  Phys.\ Lett.\ B {\bf 726}, 234 (2013).

\bibitem{Chen:2004si} 
  L.~-W.~Chen, C.~M.~Ko and B.~-A.~Li,
  Phys.\ Rev.\ Lett.\  {\bf 94}, 032701 (2005).

\bibitem{Shetty:2007zg} 
  D.~V.~Shetty, S.~J.~Yennello and G.~A.~Souliotis,
  Phys.\ Rev.\ C {\bf 76}, 024606 (2007)
  [Erratum-ibid.\ C {\bf 76}, 039902 (2007)].

\bibitem{Asakawa:1989bq} 
  M.~Asakawa and K.~Yazaki,
  Nucl.\ Phys.\ A {\bf 504}, 668 (1989).

\bibitem{Kitazawa:2002bc} 
  M.~Kitazawa, T.~Koide, T.~Kunihiro and Y.~Nemoto,
  Prog.\ Theor.\ Phys.\  {\bf 108}, 929 (2002).

\bibitem{Fukushima:2008is} 
  K.~Fukushima,
  Phys.\ Rev.\ D {\bf 78}, 114019 (2008).

\bibitem{Zhang:2009mk} 
  Z.~Zhang and T.~Kunihiro,
  Phys.\ Rev.\ D {\bf 80}, 014015 (2009).

\bibitem{Bratovic:2012qs} 
  N.~M.~Bratovic, T.~Hatsuda and W.~Weise,
  Phys.\ Lett.\ B {\bf 719}, 131 (2013).

\bibitem{Harada:2011aa} 
  M.~Harada, S.~Nakamura and S.~Takemoto,
  Phys.\ Rev.\ D {\bf 86}, 021901 (2012).

\bibitem{He:2013gta} 
  B.~-R.~He and M.~Harada,
  Phys.\ Rev.\ D {\bf 88}, no. 9, 095007 (2013).

\bibitem{Jo:2009xr} 
  K.~Jo, B.~-H.~Lee, C.~Park and S.~-J.~Sin,
  JHEP {\bf 1006}, 022 (2010).


\end{thebibliography}
\end{document}